\documentclass[english,reprint,aps,prb,superscriptaddress]{revtex4-1}
\usepackage[T1]{fontenc}
\usepackage[latin9]{inputenc}
\usepackage{amsmath}
\usepackage{amssymb}
\usepackage{graphicx}

\makeatletter
\usepackage{color}
\usepackage{hyperref}
\hypersetup{
  colorlinks=true,
  citecolor=blue,
  linkcolor=blue,
  urlcolor=blue,
}

\usepackage{times}
\usepackage[small,raggedright]{titlesec}
\titleformat*{\section} {\small\bf\uppercase}
\titlespacing*{\section} {0pt}{10pt}{0pt}

\makeatother

\usepackage{babel}
\begin{document}
\title{Steady-state Peierls transition in nanotube quantum simulator}
\author{Lin Zhang}
\email{lin.zhang@icfo.eu}
\affiliation{
ICFO-Institut de Ciencies Fotoniques, The Barcelona Institute of
Science and Technology, Castelldefels (Barcelona) 08860, Spain}
\author{Utso Bhattacharya}
\affiliation{
ICFO-Institut de Ciencies Fotoniques, The Barcelona Institute of
Science and Technology, Castelldefels (Barcelona) 08860, Spain}
\author{Adrian Bachtold}
\affiliation{
ICFO-Institut de Ciencies Fotoniques, The Barcelona Institute of
Science and Technology, Castelldefels (Barcelona) 08860, Spain}
\author{Stefan Forstner}
\affiliation{
ICFO-Institut de Ciencies Fotoniques, The Barcelona Institute of
Science and Technology, Castelldefels (Barcelona) 08860, Spain}
\author{Maciej Lewenstein}
\affiliation{
ICFO-Institut de Ciencies Fotoniques, The Barcelona Institute of
Science and Technology, Castelldefels (Barcelona) 08860, Spain}
\affiliation{ICREA, Pg. Lluis Companys 23, 08010 Barcelona, Spain}
\author{Fabio Pistolesi}
\affiliation{Univ. Bordeaux, CNRS, LOMA, UMR 5798, F-33400 Talence, France}
\author{Tobias Grass}
\affiliation{
ICFO-Institut de Ciencies Fotoniques, The Barcelona Institute of
Science and Technology, Castelldefels (Barcelona) 08860, Spain}
\affiliation{
  DIPC - Donostia International Physics Center, Paseo Manuel de Lardiz{\'a}bal 4, 20018 San Sebasti{\'a}n, Spain
}
\affiliation{
  Ikerbasque - Basque Foundation for Science, Maria Diaz de Haro 3, 48013 Bilbao, Spain
}

\begin{abstract}
Quantum dots placed along a vibrating nanotube provide a {quantum simulation platform} that can directly address the electron-phonon interaction. This offers promising prospects for the search of new quantum materials and the study of strong correlation effects. As this platform is naturally operated by coupling the dots to an electronic reservoir, state preparation is straightforwardly achieved by driving into the steady state. Here we show that for intermediate electron-phonon coupling strength, the system with spin-polarized quantum dots undergoes a Peierls transition into an insulating regime which exhibits charge-density wave order in the steady state as a consequence of the competition between electronic Coulomb repulsive interactions and phonon-induced attractive interactions. The transport phenomena can be directly observed as fingerprints of electronic correlations. We also present {powerful methods} to numerically capture the physics of such an open electron-phonon system at large numbers of phonons. Our work paves the way to study and detect {correlated electron-phonon physics} in the nanotube quantum simulator with current experimentally accessible techniques.
\end{abstract}
\maketitle

\section*{Introduction}

\begin{figure*}
\includegraphics[width=0.9\textwidth]{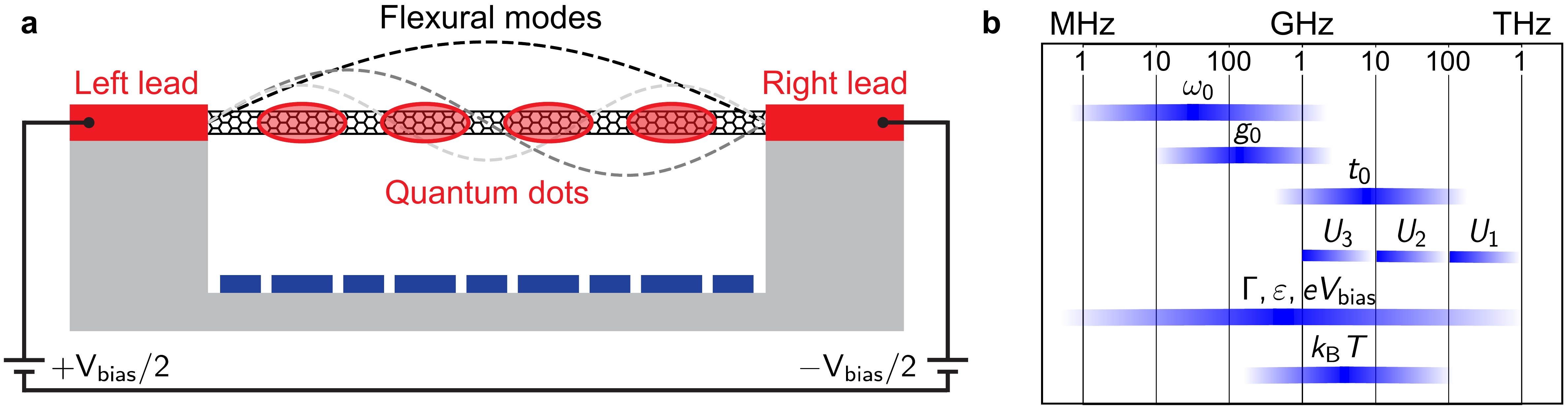}
\caption{{\bf Setup and tuning range of parameters.} \textbf{a} Schematic picture of the system. A suspended nanotube hosts four spin-polarized quantum dots. The back gates control the local chemical potential and hoppings between dots. The short separation between the nanotube and gates allows one to reach a very large electron-phonon coupling strength. The whole system couples to the left and right leads controlled by a bias voltage $V_{\mathrm{bias}}$. \textbf{b} The typical tuning range of different parameters. Here the fundamental frequency $\omega_{0}$ is divided by $2\pi$, and other parameters are divided by $2\pi\hbar$. \label{fig:figure1}}
\end{figure*}

In the search for novel materials, the simulation of quantum matter is an extremely demanding computational task which is expected to profit substantially from the surge of quantum technologies. Quantum algorithms for programmable quantum computers offer the most flexible approaches~\cite{Bauer2020,King2021,Bassman2021}, but tailor-made quantum simulation devices are particularly well suited for large-scale simulations. For instance, tremendous efforts have been made to simulate graphene-like materials using synthetic quantum systems~\cite{Polini2013}, such as artificial semiconductor lattices~\cite{Singha11}, cold atoms in optical lattices~\cite{Soltan-Panahi2011,Tarruell2012}, or photonic crystals~\cite{Plotnik2014}. However, in order to reach the complexity of real materials, quantum simulators need more controllable degrees of freedom. Specifically, apart from electrons, also phonons are fundamental ingredients of quantum materials, and the ubiquitous electron-phonon interaction is a central issue in condensed matter and material physics~\cite{Mahan2000,Ziman2001,Giustino2017}. Many important and interesting phenomena are rooted in this interaction, dating back to the seminal work on Bardeen-Cooper-Schrieffer superconductivity~\cite{bardeen57}, or charge-density wave (CDW) order induced by the Peierls instability~\cite{Peierls}. 

An archetypal and simplified Hamiltonian with the electron-electron and electron-phonon interactions is the so-called Hubbard-Holstein model, which has been studied vastly~\cite{berger_two-dimensional_1995,Ning2006,Matsueda2006,Hohenadler2013,Weber2018,Ohgoe2017,Nowadnick2012,Jansen2021}. Although many numerical and analytical approaches have been employed to solve this model, including quantum Monte Carlo~\cite{clay2005}, density-matrix renormalization group (DMRG)~\cite{fehske2008}, variational ansatz~\cite{alder1997,wang2020}, dynamical mean-field theory (DMFT)~\cite{werner2007},  and density-matrix embedding theory (DMET)~\cite{sandhoefer2016}, it is still hard to study the strongly coupled electron-phonon systems for the large number of phonons. Approaching the electron-phonon models via quantum simulation and quantum computation techniques is challenging as well. With phonons being essentially absent in optical lattices, atomic quantum simulators rely on the explicit construction of dynamical lattices as recently proposed in refs.~\cite{PhysRevLett.111.080501,PhysRevLett.121.090402} or in the context of simulation of lattice gauge theory (cf. refs.~\cite{Banyuls,Aidelsburger}), characterized by dynamical degrees of freedom on the bonds of the lattice. On the other hand, phonons occur as a natural ingredient in trapped ion quantum simulators. These systems are based on engineering the spin-phonon interactions, and hence appear to be quite natural candidates for the simulation of Holstein models~\cite{Mezzacapo2012,Knoerzer2022}. However, mapping the trapped ions system onto electron-phonon problems via Jordan-Wigner transformation is not straightforward due to long-range couplings. Another approach could be the use of digital quantum computers. In this context, a scheme of mapping electron-phonon systems onto qubits has been proposed in ref.~\cite{Macridin2018}.

On the other hand, there are also quantum simulation platforms which themselves consist of electrons and phonons and hence appear to be ideally suited for the study of quantum materials. One such platform that has recently been proposed in ref.~\cite{Bhattacharya2021} is quantum dots defined on a suspended carbon nanotube, where the phononic degrees of freedom are naturally provided by the flexural modes of the nanotube and the electrons localized in quantum dots interact with the mechanical modes electrostatically; see Fig.~\ref{fig:figure1}a for a sketch of the setup. Many parameters of the system can be tuned either at the fabrication stage or during the experiments (Fig.~\ref{fig:figure1}b). Especially, the short and controllable separation between the nanotube and gates allows one to reach a very large electron-phonon coupling strength~\cite{vigneau_ultrastrong_2021}. The nonlocal nature of phonon modes in this system also provides opportunities to explore physics beyond the Hubbard-Holstein model. Although the system is now limited in size, advances in nanofabrication make it promising to fabricate a carbon natotube with many quantum dots. On the other hand, even with the current small system, there are many interesting physical and technological phenomena associated, such as the phonon-induced pairing~\cite{Bhattacharya2021} and the nanomechanical qubit \cite{Pistolesi2021}.

As a simultaneous challenge and opportunity, one particular feature of this system is that the quantum dot setup can barely be viewed as an isolated quantum system, but rather as an open quantum system which is coupled to a fermionic environment via leads. It would be important and experimentally relevant to be able to describe the transport and driving in this challenging system with many electronic and phononic degrees of freedom. While the open system properties of single and double dots coupled to a single phonon mode has been studied extensively~\cite{Woodside1098,knobel_nanometre-scale_2003,koch_franck-condon_2005,naik_cooling_2006,mozyrsky_intermittent_2006,pistolesi_self-consistent_2008,steele_strong_2009,lassagne_coupling_2009,leturcq_franckcondon_2009,ganzhorn_dynamics_2012,benyamini_real-space_2014,pirkkalainen_cavity_2015,micchi_mechanical_2015,ares_resonant_2016,avriller_bistability_2018,khivrich_nanomechanical_2019,wen_coherent_2020}, the case of multiple dots has never been considered to our knowledge and may support interesting physics beyond single and double dots. 

In this work, we study the correlated physics in a {nanotube quantum simulator} with four spin-polarized quantum dots coupled to the leads. We develop methods to attack the challenge of theoretically describing this electron-phonon open quantum system problem. Specifically, we generalize the shift method, developed in ref.~\cite{Bhattacharya2021} to facilitate the computational treatment of equilibrium systems with large phonon numbers, to the case of open quantum systems. This treatment then allows us to study transport through the device, a great opportunity to detect important system properties. In particular, the electric current provides an immediate smoking gun of a striking behavior of the device: Upon increasing the electron-phonon coupling strength, a Peierls instability turns the system into an insulator with alternating CDW pattern of empty and occupied quantum dots as a consequence of the competition between electronic Coulomb repulsive interactions and phonon-induced attractive interactions, which is beyond the standard Hubbard-Holstein model. Although such a CDW order is also present without coupling the system to leads, observing it as a steady state property in open quantum system provides a feasible route for the preparation as well as the detection of this intriguing phenomenon. Especially, despite the fact that the current always vanishes at low enough bias due to the finite charge gap in the considered finite-size system, the different behavior of critical bias voltage for nonvanishing current as we increase the electron-phonon coupling strength directly reflects the different nature of insulating steady state, identifying the Peierls transition. 

\section*{Results}

\noindent\textbf{System}. The central component of our system, depicted in Fig.~\ref{fig:figure1}a, is a suspended carbon nanotube which hosts up to four electrons on equally spaced quantum dots spin-polarized by a large magnetic field, interacting among themselves through long-range Coulomb interactions and with the nanotube's flexural modes. Accordingly, the system Hamiltonian
\begin{equation}
  H_{\rm S}=H_{\mathrm{e}}+H_{\mathrm{p}}+H_{\text{e-p}}
\end{equation}
consists of a Hubbard-like electronic part
\begin{equation}
    H_{\mathrm{e}} = \sum_{i} \big[ -t_{0}(d_{i}^{\dagger}d_{i+1}+\mathrm{H.c.})+\varepsilon n_{i}  + \sum_{j>i} U_{j-i}n_{i}n_{j} \big],
\end{equation}
a phononic part 
\begin{equation} 
 H_{\mathrm{p}}=\sum_{\alpha}\hbar\omega_{\alpha}b_{\alpha}^{\dagger}b_{\alpha},
\end{equation} 
and the coupling between electrons and phonons 
\begin{equation}
H_{\text{e-p}}=\sum_{i,\alpha}g_{i,\alpha}n_{i}(b_{\alpha}+b_{\alpha}^{\dagger}).
\end{equation} 
Here $d_{i}$ is the spinless annihilation operator for electron on the $i$-th quantum dot, and $n_{i}=d_{i}^{\dagger}d_{i}$ is the electron number operator. Note that a single quantum dot cannot be occupied by two electrons due to the Pauli exclusion principle. The bosonic operator $b_{\alpha}$ is the annihilation operator of the phonon mode $\alpha$ with frequency $\omega_{\alpha}=\alpha\omega_{0}$ being an integer multiple of the fundamental frequency $\omega_0$, and $\hbar$ is the reduced Planck constant. {The electrons couple to the phonons through the capacitance between the quantum dots and gates, which is modulated by the displacement of the nanotube. For the coupling strength $g_{i,\alpha}$, we assume the validity of the guitar-string model~\cite{Bhattacharya2021}, in which we expand the capacitance charge energy at small displacement in terms of different oscillating modes and integrate the coupling strength between the charge and mode displacement $b_{\alpha}+b^{\dagger}_{\alpha}$ over the dot extension (assumed to be $1/4$ of the nanotube length), yielding  $g_{i,\alpha}=g_{0}(8/\pi)\alpha^{-3/2}\sin[(2i-1)\alpha\pi/8]\sin(\alpha\pi/8)$ with an overall coupling strength $g_{0}$ that can be tuned electrostatically~\cite{Bhattacharya2021}.} We also assume that the electronic hopping $t_{0}$ is uniform and between nearest neighbors only. A local chemical potential $\varepsilon$ is included to control the particle number. The parameters $U_{j-i}$ with $j>i$ describe screened Coulomb interactions. An overview of realistic parameter ranges is illustrated in Fig.~\ref{fig:figure1}b.

For a quantum simulation of the transport behavior of the electron-phonon system, we need to take into account a fermionic environment which couples to the nanotube via two leads at the left (L) and right (R) end. The Hamiltonian of the environment reads
\begin{equation}
  H_{\mathrm{E}}=\sum_{\ell k}\varepsilon_{\ell k}c_{\ell k}^{\dagger}c_{\ell k}
\end{equation}
for $\ell=L,R$, where the electrons in leads are denoted as $c$ and are labeled by the momentum $k$. The coupling between the system and leads is given by
\begin{equation}
  H_{\rm SE}=\sum_{\ell k}t_{\ell k}d_{\ell}^{\dagger}c_{\ell k}+\mathrm{H.c.},
\end{equation}
where $d_{\ell}$ denotes a system electron in the dot at the left or right end. In the following, we consider the wide-band limit for the leads with constant density of states $\nu$. We also assume the tunneling amplitudes to be energy-independent and symmetric, i.e., $t_{\ell k}=t_{\ell}$ and $t_{L}=t_{R}$. Then the coupling strength between the system and leads can be captured by the tunneling rate $\Gamma=2\pi\nu|t_{L/R}|^{2}$.

\smallskip\noindent\textbf{Quantum master equation}. The Hamiltonian $H_{\mathrm{SE}}$ consists of four different tunneling processes between the system and leads:  (i, ii) an electron entering the system from the left or right lead; (iii, iv) an electron leaving the system to the left or right lead. It is clear that these processes occur only in a particular region of the system and depend on the occupation of the environment level $\varepsilon_{\ell k} = E_{n,i_{n}}-E_{n+1,i_{n+1}}$, where $E_{n,i_{n}}$ is the system energy for the $i_n$-th eigenstate of $H_{\rm S}$ with total electron number $n$. For systems with the tunneling rate $\Gamma$ much smaller than the temperature $k_{\mathrm{B}}T$ ($k_{\mathrm{B}}$ is the Boltzmann constant), we can use the phenomenological position and energy resolving Lindblad master equation~\cite{Kirsanskas2018} to describe the reduced density matrix $\rho$ of the system coupled to leads
\begin{align}
\hbar\frac{\partial\rho}{\partial t}=& -\mathrm{i}[H_{\rm S},\rho]
  +\sum_{\ell,\alpha=\pm}\Big(L_{\ell\alpha}\rho L_{\ell\alpha}^{\dagger} \nonumber \\ &
 -\frac{1}{2}\rho L_{\ell\alpha}^{\dagger}L_{\ell\alpha}-\frac{1}{2}L_{\ell\alpha}^{\dagger}L_{\ell\alpha}\rho\Big)
\equiv \mathcal{L}[\rho].
\label{eq:lindblad master equation}
\end{align}
Here the Lindblad operators are given by
\begin{align}
L_{\ell+}=&\sum_{n,i_{n},i_{n+1}}\sqrt{f_{\ell}(E_{n+1,i_{n+1}}-E_{n,i_{n}})\Gamma} \nonumber \\ & 
\times T_{n+1,i_{n+1};n,i_{n}}^{(\ell)}\vert E_{n+1,i_{n+1}}\rangle\langle E_{n,i_{n}}\vert
\end{align}
for processes (i, ii) and
\begin{align}
L_{\ell-}=&\sum_{n,i_{n},i_{n-1}}\sqrt{[1-f_{\ell}(E_{n,i_{n}}-E_{n-1,i_{n-1}})]\Gamma} \nonumber \\ & 
\times T_{n,i_{n};n-1,i_{n-1}}^{(\ell)*}\vert E_{n-1,i_{n-1}}\rangle\langle E_{n,i_{n}}\vert
\end{align}
for processes (iii, iv), where we have $T_{n,i_{n};n-1,i_{n-1}}^{(\ell)}=\langle E_{n,i_{n}}\vert d_{\ell}^{\dagger}\vert E_{n-1,i_{n-1}}\rangle$ after rewritting the tunneling Hamiltonian $H_{\text{SE}}$ into the eigenstate basis of $H_{\mathrm{S}}$. The Fermi-Dirac distribution $f_{\ell}(x)=[e^{(x-\mu_{\ell})/k_{\mathrm{B}}T}+1]^{-1}$ with temperature $T$ and chemical potential $\mu_{\ell}$ captures the occupation of environment levels in the leads. It is clear that the rate of adding an electron to the system should be proportional to the occupation in the leads, while it is hard to remove an electron from the system if the corresponding state has already been occupied in the leads. Here we use the bias voltage $V_{\mathrm{bias}}$ to control the chemical potentials in leads and have $\mu_{L/R}=\pm eV_{\mathrm{bias}}/2$ ($e$ is the elementary electric charge), which is an important tuning knob in the open quantum system.

\begin{figure}
  \centering
  \includegraphics{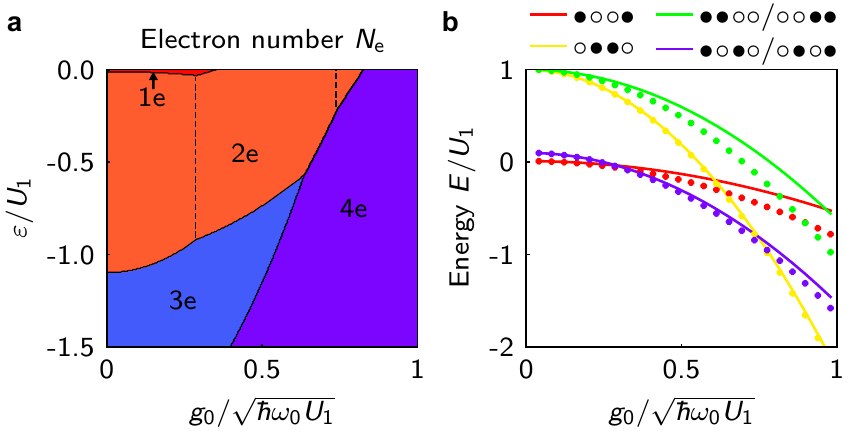}
  \caption{{\bf Role of electron-phonon coupling in the equilibrium system at atomic limit $t_{0}=0$.} \textbf{a} Electron number $N_{\rm e}$ in equilibrium as a function of local chemical potential $\varepsilon/U_1$ and electron-phonon coupling strength $g_0/\sqrt{\hbar\omega_0 U_1}$. \textbf{b} Energy of different electron configurations with fixed $N_{\rm e}=2$. Here we have used the single-mode approximation, which is verified by comparing the results obtained from single mode (solid lines) and multi (e.g., 100 considered here) modes (dots). The single- and multi-mode scenarios are found to agree with each other qualitatively well. The transition points between patterns \mbox{$\bullet\!\circ\!\circ\hspace{0.05em}\bullet$} (\mbox{$\circ\!\bullet\!\bullet\hspace{0.05em}\circ$}) and \mbox{$\bullet\!\circ\!\bullet\hspace{0.05em}\circ$}/\mbox{$\circ\!\bullet\!\circ\hspace{0.05em}\bullet$} are also presented in the two-electron region of \textbf{a} by dashed lines, respectively. Here we set $U_{1}/2\pi\hbar=200\,\mathrm{GHz}$, $U_{2}/2\pi\hbar=20\,\mathrm{GHz}$, $U_{3}/2\pi\hbar=2\,\mathrm{GHz}$, and $\omega_{0}/2\pi=3\,\mathrm{GHz}$, which are within the typical tuning range of parameters shown in Fig.~\ref{fig:figure1}b.
\label{fig:figure2}}
\end{figure}

The Lindblad master equation describes the evolution into the steady state $\rho_{\mathrm{ss}}$, which is obtained in the infinite time limit, i.e. for $t\to\infty$, or by diagonalizing the Liouvillian superoperator $\mathcal{L}$. A challenge to solve this problem is how to treat the phononic degrees of freedom for its infinite-dimensional nature. Even after truncating the phononic Hilbert space, for realistic system parameters the phonon number remains beyond our numerical capability (limited to tens of phonons). For this, we have developed different methods which facilitate the description by shifting the phononic vacuum into a state with finite phonon number and only taking into account a small number of necessary (tilded) phononic states that are coupled to the electronic states effectively. In the following, we would first focus on the physical results and shall describe the details of shift method in the ``Methods'' section for interested readers.

\begin{figure*}
  \includegraphics{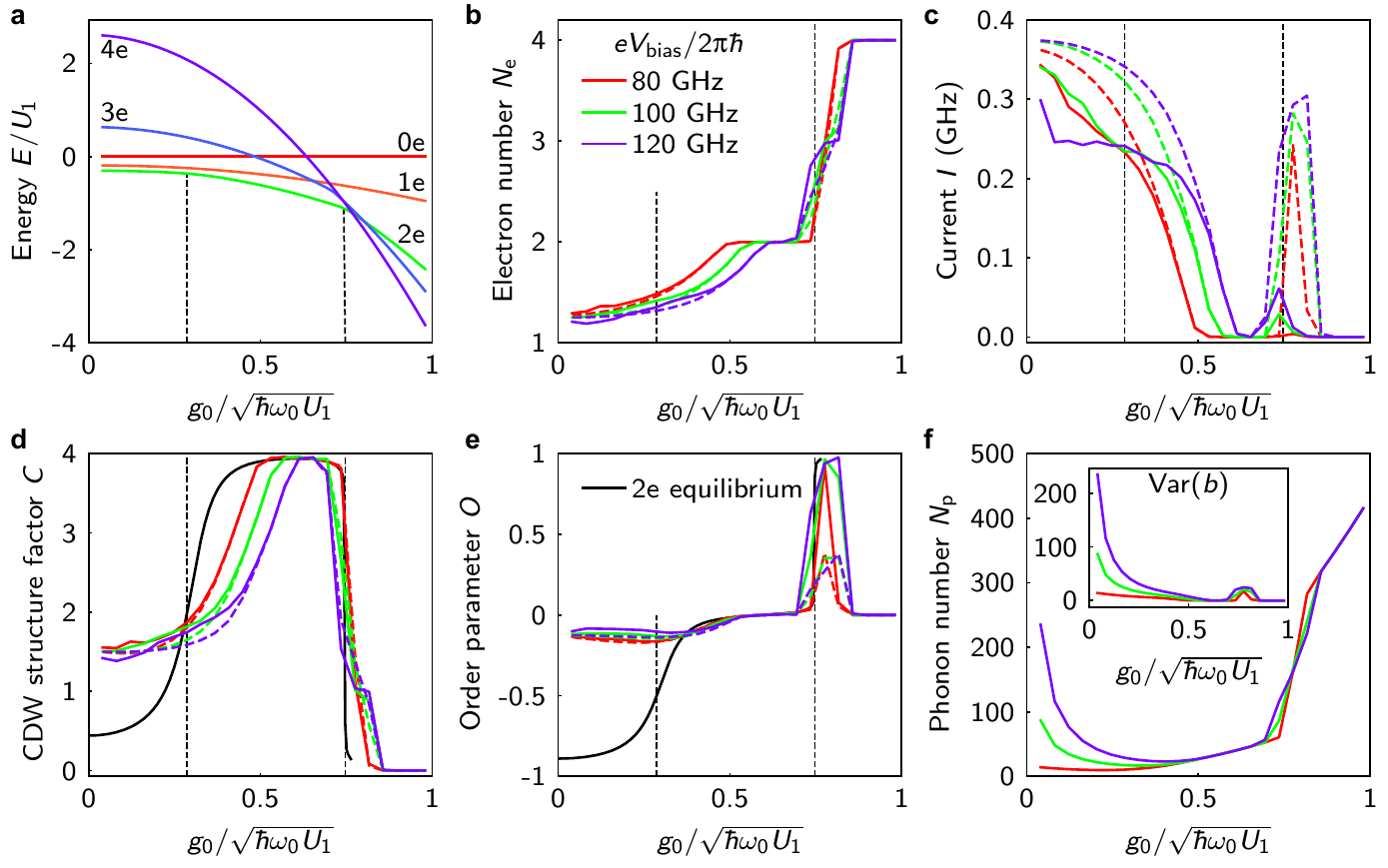}
  \caption{{\bf Phonon-induced transport behavior.} \textbf{a} Lowest energies of the system in different electron number sectors. Upon increasing the electron-phonon coupling $g_{0}$, the electron number of ground state changes from two electrons to four electrons for the chosen local chemical potential $\varepsilon/2\pi\hbar=-30\,\mathrm{GHz}$. \textbf{b}-\textbf{e} Electron number $N_{\mathrm{e}}$, current $I$, CDW structure factor $C$, and order parameter $O$ in the steady state for various $V_{\mathrm{bias}}$ obtained from the Pauli master equation (dashed lines) and shift method without updating the shift parameter (solid lines); see Methods. In \textbf{d} and \textbf{e}, we also plot the CDW structure factor $C$ and order parameter $O$ of equilibrium two-electron states (black lines) for the convenience of comparison. The black dashed vertical lines in \textbf{a}-\textbf{e} label the approximate transition points in the two-electron sector identified from the equilibrium CDW structure factor and order parameter. \textbf{f} Phonon number $N_{\mathrm{p}}$ obtained from the Pauli master equation. The insert shows the variance of phonon operator, $\mathrm{Var}(b)$. Here we truncate the phononic Hilbert space to a finite-dimensional Hilbert space with maximal 1000 phonons for the Pauli master equation.  The shift method for $eV_{\mathrm{bias}}/2\pi\hbar=80$, $100$, and $120\,\mathrm{GHz}$ is performed within a Hilbert space with maximal $40$, $45$, and $50$ tilded phonons, respectively. Other parameters are $t_{0}/2\pi\hbar=5\,\mathrm{GHz}$, $U_{1}/2\pi\hbar=200\,\mathrm{GHz}$, $U_{2}/2\pi\hbar=20\,\mathrm{GHz}$, $U_{3}/2\pi\hbar=2\,\mathrm{GHz}$, $\omega_{0}/2\pi=3\,\mathrm{GHz}$, $\Gamma/2\pi\hbar=1\,\mathrm{GHz}$, and $k_{\mathrm{B}}T/2\pi\hbar=2\,\mathrm{GHz}$.\label{fig:figure3}}
\end{figure*}

We would like to mention that if the tunneling rate $\Gamma$ is much smaller than the temperature $k_{\mathrm{B}}T$ and the energy splittings between states with the same number of electrons, a convenient simplification to the Lindblad master equation can be achieved by ignoring coherences~\cite{Schultz2009,Goldozian2016}, and we only need to consider the diagonal terms of the density matrix, $\rho_{n,i_{n}}\equiv\rho_{n,i_{n};n,i_{n}}$. This approximation leads to the Pauli master equation
\begin{equation}\label{eq:pauli master equation}
\begin{aligned}
\hbar\frac{\partial}{\partial t}\rho_{n,i_{n}}= & \sum_{\ell}\sum_{i_{n\pm1}}|[L_{\ell\mp}]_{n,i_{n};n\pm1,i_{n\pm1}}|^{2}\rho_{n\pm1,i_{n\pm1}}\\
 & -\sum_{\ell}\sum_{i_{n\pm1}}|[L_{\ell\pm}]_{n\pm1,i_{n\pm1};n,i_{n}}|^{2}\rho_{n,i_{n}}.
\end{aligned}
\end{equation}
Although this approximation is not always valid, we can use the phonon number from Pauli master equation as a good starting point for the shift method; see Methods.

\smallskip\noindent\textbf{Role of electron-phonon coupling in equilibrium}. The main application of our quantum simulation platform is to study the correlated electron-phonon physics. Hence we focus on the role played by the tunable electron-phonon coupling. Before turning our attention to the phonon-induced transport behavior in the steady state, we will first briefly show the role of electron-phonon coupling in the equilibrium system. 

The role of electron-phonon coupling in equilibrium can be most easily observed in the case of $t_{0}\to 0$. In this atomic limit, the electron-phonon problem can be solved analytically via the Lang-Firsov transformation $U=e^{-\sum_{i,\alpha}g_{i,\alpha}n_{i}(b_{\alpha}^{\dagger}-b_{\alpha})/\hbar\omega_{\alpha}}$, which yields the effective system Hamiltonian
\begin{equation}\label{eq:Lang-Firsov transformation}
  \tilde{H}_{\rm S}=H_{\mathrm{e}}\vert_{t_{0}=0}+H_{\mathrm{p}}-\sum_{\alpha}\frac{1}{\hbar\omega_{\alpha}}\left(\sum_{i}g_{i,\alpha}n_{i}\right)^{2}.
\end{equation}
Here the electron-phonon coupling is replaced by an effective phonon-induced long-range electron-electron interaction. The first role played by electron-phonon coupling is that the phonons mediate an attractive interaction, thus lowering the energy of states with more electrons; see Fig.~\ref{fig:figure2}a for the electron number $N_{\mathrm{e}}=\langle\sum_{i}n_{i}\rangle$ in equilibrium. Second, even at a fixed electron number, the electron-phonon coupling has a strong effect on the distribution of electrons along the quantum dots, as shown in Fig.~\ref{fig:figure2}b for $N_{\rm e}=2$. The phonon-induced attractive electron-electron interaction selects the two inner dots \mbox{$\circ\!\bullet\!\bullet\hspace{0.05em}\circ$}, which directly competes with the repulsive Coulomb interactions that favor the two-electron state \mbox{$\bullet\!\circ\!\circ\hspace{0.05em}\bullet$} occupying the two outer dots. Interestingly, an alternating CDW pattern \mbox{$\bullet\!\circ\!\bullet\hspace{0.05em}\circ$}/\mbox{$\circ\!\bullet\!\circ\hspace{0.05em}\bullet$} of empty and occupied dots occurs as a compromise between the two interactions for intermediate values of electron-phonon coupling.

A few remarks are in order. First, it is clear from the Lang-Firsov transformation that the phonon-induced electron-electron interaction is proportional to $g^{2}_{0}/\hbar\omega_{0}$. For a large fundamental frequency, we also need a large electron-phonon coupling to obtain the same phonon-induced electron-electron interaction strength, but the underlying physics remains the same in different parameter regime. For this and to avoid a too large phononic Hilbert space for open quantum systems, we set $\omega_{0}/2\pi=3\,\mathrm{GHz}$ for most of our calculations, which is slightly larger than the typical value in experiments but is still reasonable; see Fig.~\ref{fig:figure1}b. Second, since the phonon-induced interaction decays in the scaling of $\mathcal{O}(\alpha^{-4})$ for higher phonon modes, the higher modes would only affect the physics quantitatively but {\it not} qualitatively; see Fig.~\ref{fig:figure2}b for a comparison between the single-mode approximation and multi-mode scenario. Hence it is a good approximation to consider only the lowest phonon mode without losing the essential physics. In the following studies, we will always take this approximation and omit the subscript index of the lowest phonon mode (i.e., $b=b_{1}$) unless stated otherwise. Third, we note that the three equilibrium regimes discovered above are not limited to the atomic limit but still survive for a small hopping coefficient $t_{0}$; see Fig.~\ref{fig:figure3} and Fig.~\ref{fig:figure5}. The finite hopping problem is solved via the equilibrium shift method proposed in ref.~\cite[]{Bhattacharya2021}; see also Methods for a review. We fix the hopping coefficient as $t_{0}/2\pi\hbar=5\,\mathrm{GHz}$ in the following studies of open quantum systems. For this value, the above discussed regimes are still clearly present.

\begin{figure}
  \centering
  \includegraphics{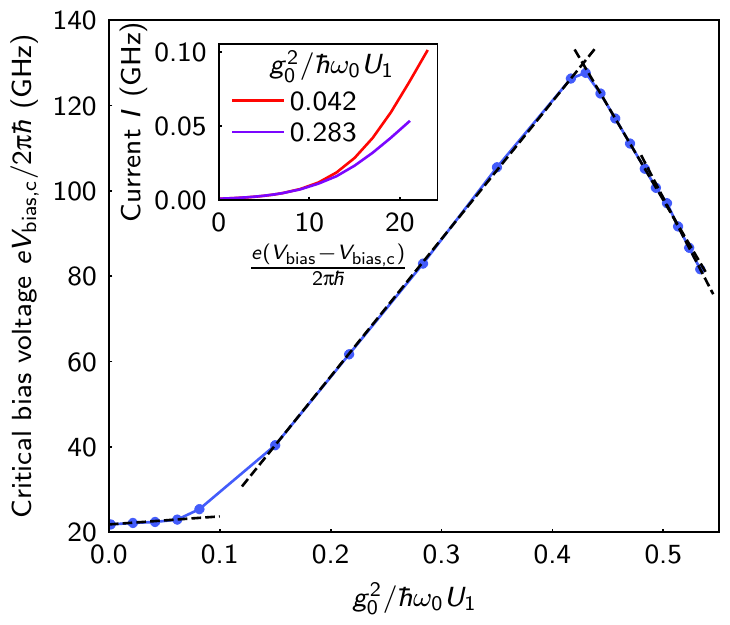}
  \caption{{\bf Critical bias voltage $V_{\mathrm{bias,c}}$ for different electron-phonon coupling strength.} Below the critical bias voltage, the electric current is smaller than $10^{-3}\,\mathrm{GHz}$, a very small value (we show two examples of the current-voltage curves in the insert). The black dashed lines are linear fittings of critical bias voltage as a function of $g^{2}_{0}/\hbar\omega_{0}U_{1}$ for different electron-phonon coupling regions. Other parameters are the same as in Fig.~\ref{fig:figure3}.
\label{fig:figure4}}
\end{figure}

\smallskip\noindent\textbf{Phonon-induced transport behavior}. With the above insight for the equilibrium properties of the system, we now characterize the steady state. The results are presented in Fig.~\ref{fig:figure3}. Since the local chemical potential $\varepsilon$ also influences the electron population (cf. Fig.~\ref{fig:figure2}a) and the interesting physics mainly occurs in the regime with $N_{\mathrm{e}}=2$, here we choose a local chemical potential such that the equilibrium state with two electrons starts from $g_{0}=0$. In Fig.~\ref{fig:figure3}a, we plot the lowest energies of different electron number sectors for the chosen local chemical potential $\varepsilon/2\pi\hbar=-30\,\mathrm{GHz}$. The equilibrium ground state changes from the two-electron sector to the four-electron sector upon increasing the electron-phonon coupling. The range of electron-phonon coupling strength with two electrons is also large for this local chemical potential, where different two-electron regimes can happen, although the range for four electrons is even larger and goes to infinity. We note that since the region for equilibrium state \mbox{$\circ\!\bullet\!\bullet\hspace{0.05em}\circ$} is quite small compared to other two-electron regimes and is very close to the four-electron state for the local chemical potentials with two-electron states starting from $g_{0}=0$ in equilibrium (see Fig.~\ref{fig:figure2}a), we mainly focus on the nonequilibrium properties of patterns \mbox{$\bullet\!\circ\!\circ\hspace{0.05em}\bullet$} and \mbox{$\bullet\!\circ\!\bullet\hspace{0.05em}\circ$}/\mbox{$\circ\!\bullet\!\circ\hspace{0.05em}\bullet$} in the following studies.

The electronic transport properties in steady state for various bias voltage $V_{\mathrm{bias}}$ are shown in Figs.~\ref{fig:figure3}b-\ref{fig:figure3}e, where both the results from Pauli master equation and the generalized shift method for open electron-phonon systems (see Methods) are provided and match with each other qualitatively. For the electron number $N_{\mathrm{e}}=\mathrm{Tr}[\rho_{\mathrm{ss}}\sum_{i}n_{i}]$, we observe a similar behavior like the one presented in equilibrium regime that upon increasing the electron-phonon coupling, the number of electrons in the steady state is increased; see Fig.~\ref{fig:figure3}b. But now the electron numbers are not quantized in general due to the mixing of eigenstates in different electron number sectors. Strikingly, we also find plateaus with electron numbers $N_{\rm e}\approx 2$ and $N_{\rm e}\approx 4$ in the steady state for a moderate bias voltage. Since it is necessary that the number of electrons fluctuate over time in order to have sequential transport, a constant and integer value of $N_{\mathrm{e}}$ signals a blockade of the steady state current $I=I_{L}=-I_{R}$ defined by
\begin{equation}
  I_{\ell}=(1/2\pi\hbar)\mathrm{Tr}[\rho_{\mathrm{ss}}(L_{\ell+}^{\dagger}L_{\ell+}-L_{\ell-}^{\dagger}L_{\ell-})],
\end{equation}
which characterizes the net rate of electron number flow through the system and vanishes in these regions; see Fig.~\ref{fig:figure3}c.

While the insulating nature of four-electron state is a trivial consequence of Pauli blocking, it is a very intriguing behavior for $N_{\rm e}=2$. In fact, this insulating regime coincides with a CDW pattern in the electronic configuration. To detect this pattern, we introduce the structure factor
\begin{equation}
  C=\sum_{i,j}(-1)^{i-j}\mathrm{Tr}[\rho_{\mathrm{ss}}n_{i}n_{j}],
\end{equation}
which takes the value $C=4$ in the perfect CDW order. Indeed, the $N_{\rm e}=2$ plateau in Fig.~\ref{fig:figure3}b coincides with CDW order; see Fig.~\ref{fig:figure3}d. As indicated by the black line, this regime also exhibits CDW order in equilibrium. Therefore, this pattern is not induced by the tunneling to leads but indeed by the competition between repulsive Coulomb interaction and phonon-induced attractive electron-electron interaction. 

We also introduce the order parameter
\begin{equation}
  O=\mathrm{Tr}[\rho_{\mathrm{ss}}n_{2}n_{3}]-\mathrm{Tr}[\rho_{\mathrm{ss}}n_{1}n_{4}]
\end{equation}
to characterize the electron configurations \mbox{$\circ\!\bullet\!\bullet\hspace{0.05em}\circ$}/\mbox{$\bullet\!\circ\!\circ\hspace{0.05em}\bullet$} through values $O=\pm1$, and $O=0$ for the CDW pattern. The steady-state values of $O$ are plotted in Fig.~\ref{fig:figure3}e, together with the equilibrium value of two-electron states. For small electron-phonon coupling, while the equilibrium system exhibits the pattern \mbox{$\bullet\!\circ\!\circ\hspace{0.05em}\bullet$}, the order parameter $O$ is highly reduced in the nonequilibrium steady state due to the enhanced mixing between patterns \mbox{$\bullet\!\circ\!\circ\hspace{0.05em}\bullet$} and \mbox{$\bullet\!\circ\!\bullet\hspace{0.05em}\circ$}/\mbox{$\circ\!\bullet\!\circ\hspace{0.05em}\bullet$} and with the one-electron eigenstates. Hence it is hard to distinguish the weakly coupled regime from the CDW order via the order parameter $O$, unlike the CDW structure factor $C$.

\begin{figure*}[t]
  \includegraphics{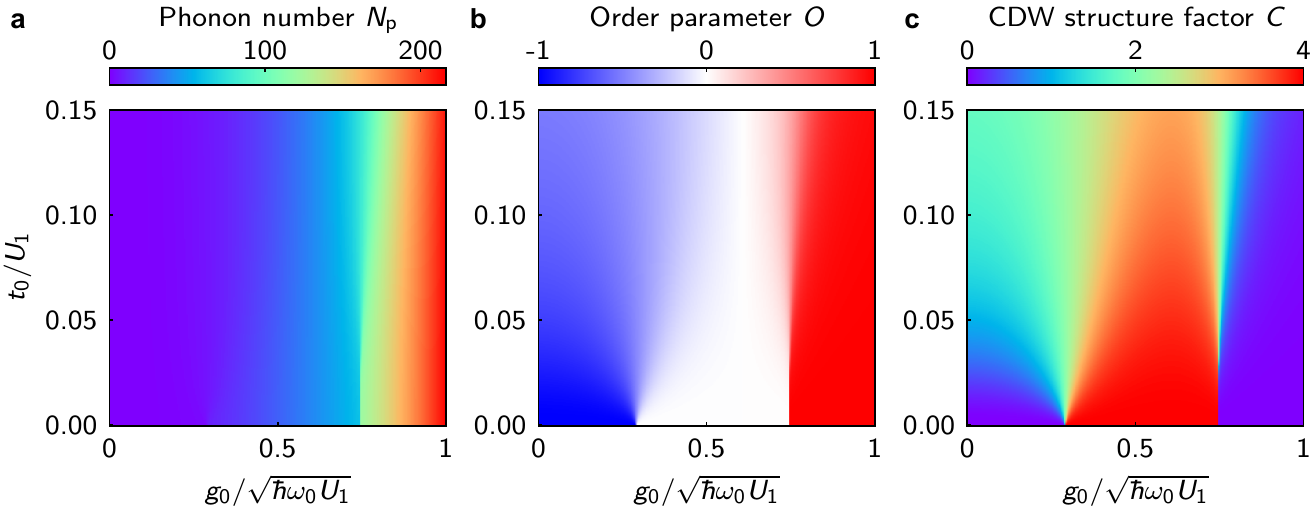}
  \caption{{\bf Equilibrium two-electron states at finite hopping $t_{0}$.} \textbf{a} The phonon number $N_{\mathrm{p}}$ increases as the electron-phonon coupling becomes stronger and is more smooth for a larger $t_{0}$. \textbf{b}, \textbf{c} The order parameter $O$ and CDW structure factor $C$ get reduced, but the three equilibrium regimes discovered in the atomic limit still survive under small but finite $t_{0}$. Here we set $U_{1}/2\pi\hbar=200\,\mathrm{GHz}$, $U_{2}/2\pi\hbar=20\,\mathrm{GHz}$, $U_{3}/2\pi\hbar=2\,\mathrm{GHz}$, and $\omega_{0}/2\pi=3\,\mathrm{GHz}$. The local chemical potential $\varepsilon$ is ignored for the two-electron case, and the bias voltage is set to zero for these plots. \label{fig:figure5}}
\end{figure*}

We note that the current is always small at low enough bias due to the finite charge gap in the considered finite-size system. To further identify the nature of insulating steady state at certain coupling strength, we define the critical bias voltage $V_{\mathrm{bias,c}}$, at which the corresponding current is $10^{-3}\,\mathrm{GHz}$, a very small value for the used tunneling rate. Since the critical bias voltage is directly related to the charge gap and the changes of energy spectra as we increase the electron-phonon coupling are given by $(\sum_{i}g_{i,1}n_{i})^{2}/\hbar\omega_{0}$ in the atomic limit [cf. Eq.~\eqref{eq:Lang-Firsov transformation}], we would expect that the critical bias voltage is proportional to $g^{2}_{0}$ and depends on the electron distributions of the coupled states by leads. Indeed, this is also confirmed for a finite but small hopping $t_{0}$ in Fig.~\ref{fig:figure4}, where the critical bias voltages can be captured by a series of linear fittings. The different slope manifests the distinct nature of the corresponding insulating steady states. 

Initially, the two-electron state in equilibrium mainly has the pattern \mbox{$\bullet\!\circ\!\circ\hspace{0.05em}\bullet$} but is also perturbed by the CDW order due to the finite $t_{0}$, which will be coupled to the one-electron states made of components \mbox{$\bullet\!\circ\!\circ\hspace{0.05em}\circ$}/\mbox{$\circ\!\circ\!\circ\hspace{0.05em}\bullet$} and \mbox{$\circ\!\bullet\!\circ\hspace{0.05em}\circ$}/\mbox{$\circ\!\circ\!\bullet\hspace{0.05em}\circ$} by leads (the coupling to three-electron states are highly suppressed by the energy gap; cf. Fig.~\ref{fig:figure3}a). Since the one-electron state dominated by pattern \mbox{$\circ\!\bullet\!\circ\hspace{0.05em}\circ$}/\mbox{$\circ\!\circ\!\bullet\hspace{0.05em}\circ$} has lower energy, the above two-electron state is mainly coupled to this state for small bias voltage, determining the charge gap. The slight $g_{0}$-dependence of energies of these states gives a small slope for the critical bias voltage. As we increase the electron-phonon coupling, the Peierls instability from pattern \mbox{$\bullet\!\circ\!\circ\hspace{0.05em}\bullet$} into the CDW order suddenly enhances the reduction of energy for two-electron state (cf. Fig.~\ref{fig:figure2}b), which is still mainly coupled to the one-electron state dominated by pattern \mbox{$\circ\!\bullet\!\circ\hspace{0.05em}\circ$}/\mbox{$\circ\!\circ\!\bullet\hspace{0.05em}\circ$}, leading to the quick increase in critical bias voltage and a large slope. This feature provides an immediate smoking gun for the Peierls transition. As the coupling further increases, the energy of three-electron state would be lower than that of the one-electron state. The two-electron state now mainly couples to the three-electron state dominated by pattern \mbox{$\bullet\!\circ\!\bullet\hspace{0.05em}\bullet$}/\mbox{$\bullet\!\bullet\!\circ\hspace{0.05em}\bullet$} (the weight of pattern \mbox{$\circ\!\bullet\!\bullet\hspace{0.05em}\bullet$}/\mbox{$\bullet\!\bullet\!\bullet\hspace{0.05em}\circ$} will increase for larger $g_{0}$ due to the reduced energy), and the charge gap as well as the critical bias voltage start to decrease, although the insulating steady state still exhibit CDW order (Fig.~\ref{fig:figure3}d). The further reduced charge gap may even lead to a small increase in the current for a fixed large bias voltage; cf. Fig.~\ref{fig:figure3}c.

Finally, we consider the phonon number $N_{\mathrm{p}}=\mathrm{Tr}[\rho_{\mathrm{ss}}b^{\dagger}b]$ in the steady state using Pauli master equation; see Fig.~\ref{fig:figure3}f {and discussions in Methods}. An abrupt increase of $N_{\mathrm{p}}$ is observed when the electron-phonon coupling becomes too strong to support the CDW structure, with little dependence on the applied bias voltage. Interestingly, for relatively large bias voltage, the phonon number also takes a large value for small electron-phonon coupling. This counter intuitive phenomenon (cf. Fig.~\ref{fig:figure5}a for phonon number in equilibrium) indeed is very different from the large phonon number at strong electron-phonon coupling.
Actually the phonon number $N_{\mathrm{p}}$ has two contributions, one coming from the polaronic displacement of the oscillator $\Delta x=\mathrm{Tr}[\rho_{\mathrm{ss}}b]+\mathrm{Tr}[\rho_{\mathrm{ss}}b^{\dagger}]$, thus $N_{\mathrm{p}} \approx (\Delta x/2)^2$,
and another from the distribution of phonons. The last part is related to the variance of phonon operator 
$\mathrm{Var}(b)= \mathrm{Tr}[\rho_{\mathrm{ss}}b^{\dagger}b]- \mathrm{Tr}[\rho_{\mathrm{ss}}b^{\dagger}]\mathrm{Tr}[\rho_{\mathrm{ss}}b]$
(see the insert of Fig.~\ref{fig:figure3}f), and measures the number of excited phonons in the system. The results indicate that for small coupling the system heats up. This is due to the current passing through the device, as one sees that in coincidence with the degeneracy point ($g_{0}/\sqrt{\hbar\omega_{0}U_{1}}\approx 0.77$) the variance has also a maximum and that 
the variance increases with the voltage in these regions. For strong electron-phonon coupling, the variance of phonon operator is quite small. This suggests that the large phonon number in this regime is only due to the polaronic displacement of the oscillator that is proportional to the electro-mechanical coupling constant $g_{0}$. The system is actually in its ground state, since the current is blocked and cannot transfer heat to the oscillator.

{We can also understand the large phonon number at small coupling from the perspective of the Pauli master equation. In the limit of $g_{0}\to 0$, the electronic and phononic degrees of freedom are indeed decoupled, which means that the lead-induced outgoing and incoming tunneling rates for an eigenstate of $H_{\mathrm{S}}$ in Eq.~\eqref{eq:pauli master equation} are uniform with respect to the phonon number in this limit. At small $g_{0}$, the weak electron-phonon coupling only has limited influence on the eigenstates, hence the tunneling rates are still very close and do not have significant difference in the whole range of phonon numbers. This allows for the occupation of excited states with large phonon number at small electron-phonon coupling. Obviously, a larger bias voltage would allow more eigenstates to be coupled, leading to a larger phonon number.}

\section*{Discussion}

In summary, we have investigated the steady state of a {quantum simulation platform} for correlated electron-phonon physics, where four spin-polarized quantum dots are equally placed along a suspended carbon nanotube and are coupled to a fermionic environment provided by two leads at the left and right end. Different regimes in the steady state as well as in equilibrium as a function of electron-phonon coupling strength have been explored. Particularly, for intermediate electron-phonon coupling, the Peierls instability drives the system into an insulating regime with CDW order as a consequence of the competition between electronic Coulomb repulsive interactions and phonon-induced attractive interactions. The different behavior of critical bias voltage for nonvanishing current further provides an immediate smoking gun for the different nature of insulating steady state, identifying the Peierls transition. To treat the electron-phonon open systems with large phonon numbers, we also developed a generalized shift method in the ``Methods'' section, which highly reduces the phononic degrees of freedom required to describe the problem.

Our work shall largely stimulate the experimental advances in this field. Particularly, the proposed setup can be accessed via current techniques in nanofabrication. The high tunability of different parameters allows us to directly observe and detect the predicted phonon-induced correlated physics and transport behavior. Although our work focuses on the spin-polarized quantum dots, it would be interesting to further take into account the spin and/or valley degrees of freedom for the open electron-phonon systems in the future studies, which shall bring in new kind of orders and nonequilibrium behavior.

\begin{figure}
  \includegraphics{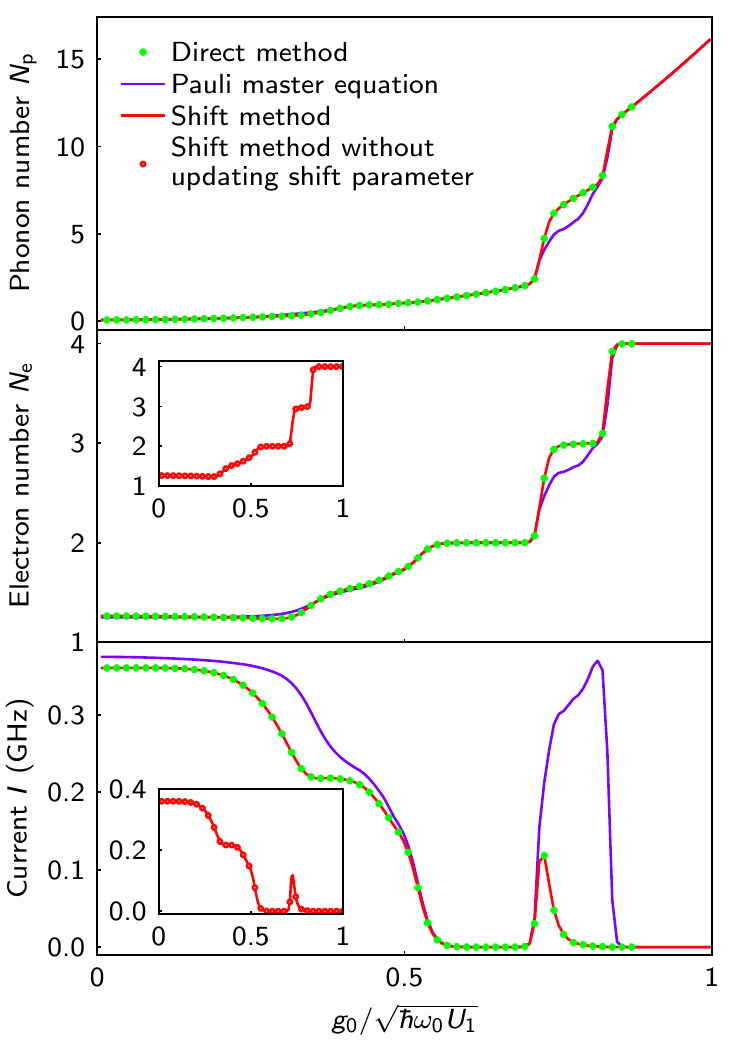}
  \caption{{\bf Comparison between different methods for solving the steady state.} Here we choose a large fundamental frequency $\omega_{0}/2\pi=80\,\mathrm{GHz}$ such that the phonon number in the system is small. We truncate the phononic Hilbert space to a finite-dimensional Hilbert space with maximal 30 (or 100) phonons for the direct method (or Pauli master equation approach). On the other hand, the shift methods with and without updating the shift parameter are performed within a Hilbert space with maximal 5 tilded phonons. Here we set $eV_{\mathrm{bias}}/2\pi\hbar=100\,\mathrm{GHz}$, and other parameters are the same as in Fig.~\ref{fig:figure3}.\label{fig:figure6}}
\end{figure}

\section*{Methods}

\noindent\textbf{Shift method for equilibrium systems at finite hopping}. In Eq.~\eqref{eq:Lang-Firsov transformation}, we have utilized the Lang-Firsov transformation to transform the equilibrium electron-phonon Hamiltonian in the atomic limit into a purely electronic problem. However, the electrons and phonons cannot be decoupled at finite hopping, since the hopping coefficient transforms as $t_{0}\to t_{0}e^{-(g_{i,1}-g_{i+1,1})(b^{\dagger}-b)/\hbar\omega_{0}}$ and will gain a phonon-dependent phase. One common approach is to truncate the infinite-dimensional phononic Hilbert space to a finite-dimensional Hilbert space and then perform the numerical diagonalization. However, this method is limited to small phonon numbers. To overcome this problem, a shift method has been developed in ref.~\cite{Bhattacharya2021}. 

The basic idea of this method is to shift the phononic vacuum to a state with finite number of phonons by the transformation $b\to\tilde{b}+S$ and only take into account a small number of effective phononic states that are coupled to the electrons (e.g., we consider a Hilbert space with maximal 30 tilded phonons). The shift parameter $S$ will be updated iteratively according to the replacement $S\leftarrow-[\langle(\tilde{b}^{\dagger}+S^{*})(\tilde{b}+S)\rangle]^{1/2}$ until convergence, where the expectation $\langle\cdot\rangle$ is performed on the ground state of the shifted Hamiltonian $H_{\mathrm{S}}[d_{i},\tilde{b},S]$. In the practical calculation, we can utilize the phonon number $N_{\mathrm{p}}=\left(\sum_{i}g_{i,1}n_{i}/\hbar\omega_{0}\right)^{2}$ in the atomic limit to obtain a good initial guess $S_{0}=-\sqrt{N_{\mathrm{p}}}$ for the shift parameter.

As an application of this method and a complement to Fig.~\ref{fig:figure2}b, we show the two-electron equilibrium states at finite hopping in Fig.~\ref{fig:figure5}, with a particular interest in the fate of CDW order at large values of hopping coefficient. First, unlike the phonon number in steady state (cf. Fig.~\ref{fig:figure3}f), the equilibrium phonon number $N_{\mathrm{p}}=\langle(\tilde{b}^{\dagger}+S^{*})(\tilde{b}+S)\rangle$ always increases as the electron-phonon coupling increases; see Fig.~\ref{fig:figure5}a. There is also a clear discontinuity for small $t_{0}$ at the transition point from CDW order to the \mbox{$\circ\!\bullet\!\bullet\hspace{0.05em}\circ$} pattern due to the first order nature of this transition, which is smoothed by large hopping coefficient. For the electronic properties, the three equilibrium regimes discovered in the atomic limit survive under small but finite $t_{0}$ and distinguish from each other clearly, as indicated in the order parameter $O$ (Fig.~\ref{fig:figure5}b) and CDW structure factor $C$ (Fig.~\ref{fig:figure5}c), although the corresponding values are reduced. On the other hand, at large hopping coefficient, the boundary between the pattern \mbox{$\bullet\!\circ\!\circ\hspace{0.05em}\bullet$} and CDW configuration is strongly blurred. This is attributed to the fact that the energy gap between these two states is small and changes slowly as we increase the electron-phonon coupling strength; cf. Fig.~\ref{fig:figure2}b. Therefore, these two states will strongly affect each other for a wide range of $g_{0}$, blurring the distinctions.

\smallskip\noindent\textbf{Generalized shift methods for steady states}. Now we generalize the shift method to open electron-phonon systems. We note that the steady state $\rho_{\mathrm{ss}}$ can be considered as the ground state of an effective Hamiltonian $\mathcal{L}^{\dagger}\mathcal{L}$. As in the shift method for equilibrium ground states, we first make the transformation $b\to\tilde{b}+S$ to shift the phononic vacuum to a state with phonon number $|S|^{2}$. Then we diagonalize the Hamiltonian $H_{\mathrm{S}}[d_{i},\tilde{b},S]$ and construct the jump operators $L_{\ell\alpha}[S]$ via the eigenstates. Finally, we solve the steady state $\rho_{\mathrm{ss}}[S]$ of the Lindblad master equation \eqref{eq:lindblad master equation} in terms of $H_{\mathrm{S}}[d_{i},\tilde{b},S]$ and $L_{\ell\alpha}[S]$ and update the shift parameter as 
\begin{equation}
S\leftarrow-\sqrt{\mathrm{Tr}\{\rho_{\mathrm{ss}}[S](\tilde{b}^{\dagger}+S^{*})(\tilde{b}+S)\}}.
\end{equation}
The above procedures are repeated until convergence of the shift parameter. It is also a good initial guess for the shift parameter to use the phonon number from Pauli master equation.

We would like to mention that the dimensionality of the shifted phononic Hilbert space in general depends on the chemical potential and temperature in the leads and the phonon frequency, which determine how many phonons will be coupled by the environment. For a large bias voltage and temperature compared with the phonon frequency, the number of tilded phonons may be still large. {In this case, although the effective phononic degrees of freedom have been highly reduced compared to the bare phonons, it is still not easy to solve the steady state. Therefore, the above method is limited to relatively small bias voltage and temperature.}

{On the other hand, we note that the fermionic environment directly couples to the electronic degrees of freedom in the system. To solve the steady state for a large tilded phononic Hilbert space, we further make an approximation for the shift method, i.e., we solve the phonon number from the Pauli master equation and use it as a shift parameter for the shift method to develop the possible coherence between electronic degrees of freedom using the time evolution or iterative method until the electronic properties converge, which is much faster than the convergence of phonon number $N_{\mathrm{p}}=\mathrm{Tr}\{\rho_{\mathrm{ss}}[S](\tilde{b}^{\dagger}+S^{*})(\tilde{b}+S)\}$. In this method, the shift parameter will not be updated, and the phonon number from Pauli master equation is assumed to be close to the real one. For the regime with small electron-phonon coupling strength or ignorable coherence, this approximation should work well, since the corresponding coherence between phononic degrees of freedom is weak. We note that even for the regime beyond this limitation, the electronic properties obtained from this method are still more reasonable than the results obtained from the Pauli master equation, as part of the coherence has been captured.}

In Fig.~\ref{fig:figure6}, we benchmark our generalized shift methods by observing the steady state phonon number $N_{\mathrm{p}}$, electron number $N_{\mathrm{e}}$, and current $I$. Here we choose a very large fundamental frequency $\omega_{0}$ such that the phonon number in the system is small and it is still possible to solve the steady state directly (i.e., without using the shift methods). Compared with the direct method by solving the equation $\mathcal{L}[\rho_{\mathrm{ss}}]=0$, the results obtained from Pauli master equation matches qualitatively. However, the difference may be large near $g_{0}/\sqrt{\hbar\omega_{0}U_{1}}\approx 0.77$, which corresponds to the equilibrium transition point from two-electron states to the four-electron states (cf. Fig.~\ref{fig:figure3}a). Here the energy levels are close to each other, and the coherence cannot be ignored. Due to the lack of coherence, the current from Pauli master equation is in general larger than the true value even if the phonon number and electron number are nearly the same. {We note that the phonon number from Pauli master equation only differs from the real one significantly in the region where the electron-phonon coupling is strong and the coherence is relevant.}

On the other hand, the results from shift method matches quantitatively with the exact results, and the effective dimension of phononic Hilbert space has been highly reduced compared with the direct method (e.g., from 30 to 5 in our calculation). Hence the shift method works quite well. We also plot the electron number and current obtained from shift method without updating the shift parameter, where the shift parameter is provided by the phonon number from Pauli master equation and we evolve the shifted Lindblad master equation until the electronic properties converge. The matching with the full shift method as well as the exact results suggests that this method is also a good approximation.

\section*{Data availability}
The data supporting the results in this work are available from the corresponding author upon reasonable request.

\section*{Code availability}
The code supporting the results in this work are available from the corresponding author upon reasonable request.


\section*{Acknowledgments}
U.B., T.G., L.Z., and M.L. acknowledge support from: ERC AdG NOQIA; Ministerio de Ciencia y Innovation Agencia Estatal de Investigaciones (PGC2018-097027-B-I00/10.13039/501100011033, CEX2019-000910-S/10.13039/501100011033, Plan National FIDEUA PID2019-106901GB-I00, FPI, QUANTERA MAQS PCI2019-111828-2, QUANTERA DYNAMITE PCI2022-132919, Proyectos de I+D+I ``Retos Colaboraci{\' o}n'' QUSPIN RTC2019-007196-7); MICIIN with funding from European Union NextGenerationEU(PRTR-C17.I1) and by Generalitat de Catalunya; Fundaci{\' o} Cellex; Fundaci{\' o} Mir-Puig; Generalitat de Catalunya (European Social Fund FEDER and CERCA program, AGAUR Grant No. 2017 SGR 134, QuantumCAT\textbackslash U16-011424, co-funded by ERDF Operational Program of Catalonia 2014-2020); Barcelona Supercomputing Center MareNostrum (FI-2022-1-0042); EU Horizon 2020 FET-OPEN OPTOlogic (Grant No 899794); EU Horizon Europe Program (Grant Agreement 101080086 -- NeQST), National Science Centre, Poland (Symfonia Grant No. 2016/20/W/ST4/00314); ICFO Internal ``QuantumGaudi'' project; European Union's Horizon 2020 research and innovation programme under the Marie-Sk\l odowska-Curie grant agreement No. 101029393 (STREDCH) and No. 847648 (``La Caixa'' Junior Leaders fellowships ID100010434: LCF/BQ/PI19/11690013, LCF/BQ/PI20/11760031, LCF/BQ/PR20/11770012, LCF/BQ/PR21/11840013).
F.P. acknowledges support from the French Agence Nationale de la Recherche (grant SINPHOCOM ANR-19-CE47-0012). 
A.B. and S.F. acknowledge ERC Advanced Grant No. 692876, MICINN Grant No. RTI2018-097953-B-I00, AGAUR (Grant No. 2017SGR1664), the Quantera grant (PCI2022-132951), the Fondo Europeo de Desarrollo, the Spanish Ministry of Economy and Competitiveness through Quantum CCAA and CEX2019-000910-S [MCIN/AEI/10.13039/501100011033], Fundacio Cellex, Fundacio Mir-Puig, Generalitat de Catalunya through CERCA. Views and opinions expressed in this work are, however, those of the author(s) only and do not necessarily reflect those of the European Union, European Climate, Infrastructure and Environment Executive Agency (CINEA), nor any other granting authority. Neither the European Union nor any granting authority can be held responsible for them.

\section*{Author contributions}
T.G. and M.L. conceived the project. L.Z., U.B. and F.P. established the master equation. A.B. and S.F. provided the typical tuning range of different parameters. L.Z. developed the generalized shift method and performed the numerical studies. All authors contributed to the analysis of numerical results and the writing of the manuscript.

\section*{Competing interests}
The authors declare no competing interests.

\end{document}